\documentclass[sn-mathphys]{sn-jnl}

\usepackage{graphicx}
\usepackage{float}
\usepackage{subfig}
\usepackage{enumerate}
\usepackage{bm}
\usepackage{multirow}
\usepackage{algorithm}
\usepackage{algpseudocode}
\usepackage{amsmath}
\usepackage{setspace}

\jyear{2021}%

\theoremstyle{thmstyleone}%
\theoremstyle{thmstyletwo}%

\theoremstyle{thmstylethree}%

\begin{document}

\title[Article Title]{SR-HetGNN: Session-based Recommendation with Heterogeneous Graph Neural Network}


\author*[1,2]{\fnm{Jinpeng} \sur{Chen}}\email{jpchen@bupt.edu.cn}

\author[1,2]{\fnm{Haiyang} \sur{Li}}
\equalcont{These authors contributed equally to this work.}

\author[1,2]{\fnm{Xudong} \sur{Zhang}}
\equalcont{These authors contributed equally to this work.}
 \author[1,2]{ Fan~Zhang}
 \author[3] {Senzhang~Wang}
 \author[4] {Kaimin~Wei}
 \author*[5,6] {\fnm{Jiaqi} \sur{Ji}}\email{52987541@qq.com}

\affil*[1]{\orgdiv{School of Computer Science (National Pilot Software Engineering School)}, \orgname{Beijing University of Posts and Telecommunications}, \orgaddress{\street{Xitucheng Road}, \city{Beijing}, \postcode{100876}, \state{Beijing}, \country{China}}}

\affil[2]{\orgdiv{Key Laboratory of Trustworthy Distributed Computing and Service (Beijing University of Posts and Telecommunications)}, \orgname{Ministry of Education}, \orgaddress{\street{Xitucheng Road}, \city{Beijing}, \postcode{100876}, \state{Beijing}, \country{China}}}

\affil[3]{\orgdiv{School of Computer Science and Engineering}, \orgname{Central South University}, \orgaddress{\street{yuelu district}, \city{Changsha}, \postcode{410083}, \state{Hunan Province}, \country{China}}}

\affil[4]{\orgdiv{College of Information Science and Technology}, \orgname{Jinan University}, \orgaddress{\street{Huangpu Avenuet}, \city{Guangzhou}, \postcode{510632}, \state{Guangdong Province}, \country{China}}}
\affil[5]{\orgdiv{The Technology Innovation Center of Cultural Tourism Big Data of Hebei Province},  \orgaddress{\city{Chengde}, \postcode{067000}, \state{Hebei Province}, \country{China}}}
\affil[6]{ \orgname{Hebei Normal University for Nationalities}, \orgaddress{ \city{Chengde}, \postcode{067000}, \state{Hebei Province}, \country{China}}}


\abstract{The Session-Based Recommendation System aims to predict the user's next click based on their previous session sequence. The current studies generally learn user preferences according to the transitions of items in the user's session sequence. However, other effective information in the session sequence, such as user profiles, are largely ignored which may lead to the model unable to learn the user's specific preferences. In this paper, we propose SR-HetGNN, a novel session recommendation method that uses a heterogeneous graph neural network (HetGNN) to learn session embeddings and capture the specific preferences of anonymous users. Specifically, SR-HetGNN first constructs heterogeneous graphs containing various types of nodes according to the session sequence, which can capture the dependencies among items, users, and sessions. Second, HetGNN captures the complex transitions between items and learns the item embeddings containing user information. Finally, local and global session embeddings are combined with the attentional network to obtain the final session embedding, considering the influence of users' long and short-term preferences. SR-HetGNN is shown to be superior to the existing state-of-the-art session-based recommendation methods through extensive experiments over two real large datasets Diginetica and Tmall.}




\maketitle

\section{Introduction}\label{sec1}

\begin{figure*}[htbp]
\centering
\subfloat[]{
\includegraphics[scale=0.33,trim=0 20 550 0,clip ]{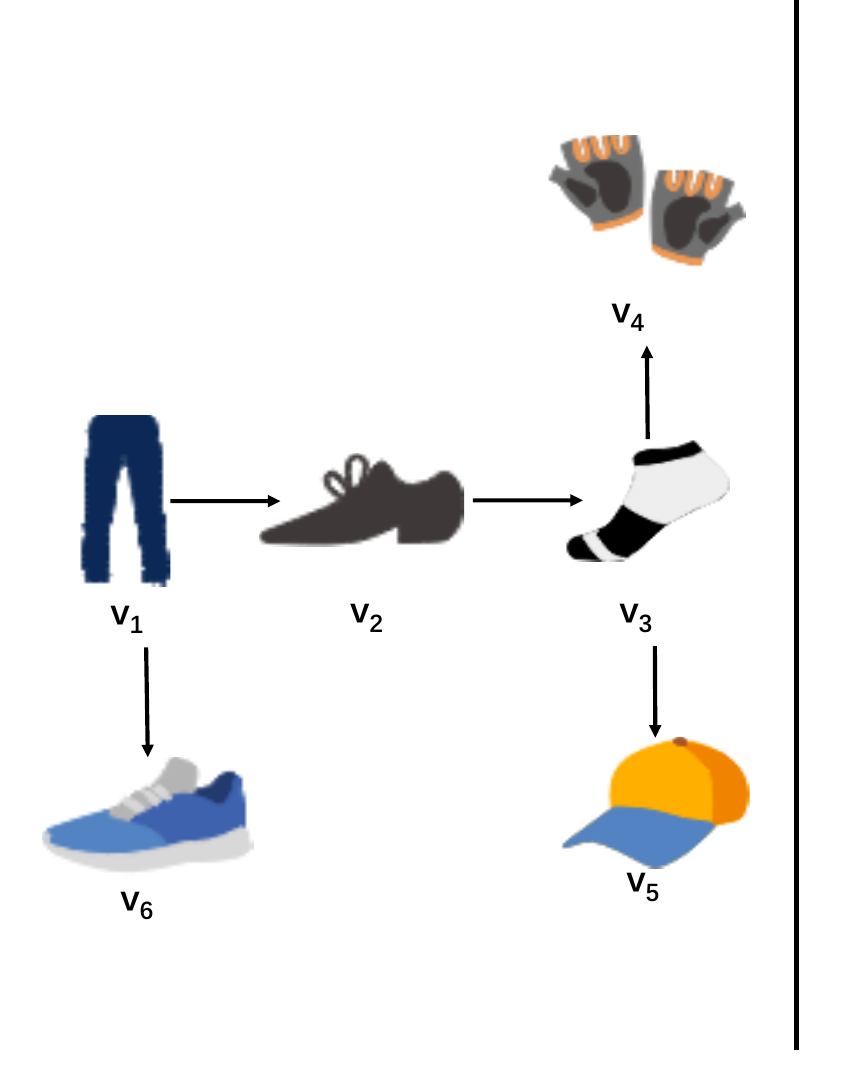}
\label{fig:1a}
}
\subfloat[]{
\includegraphics[scale=0.33,trim=420 20 0 0,clip ]{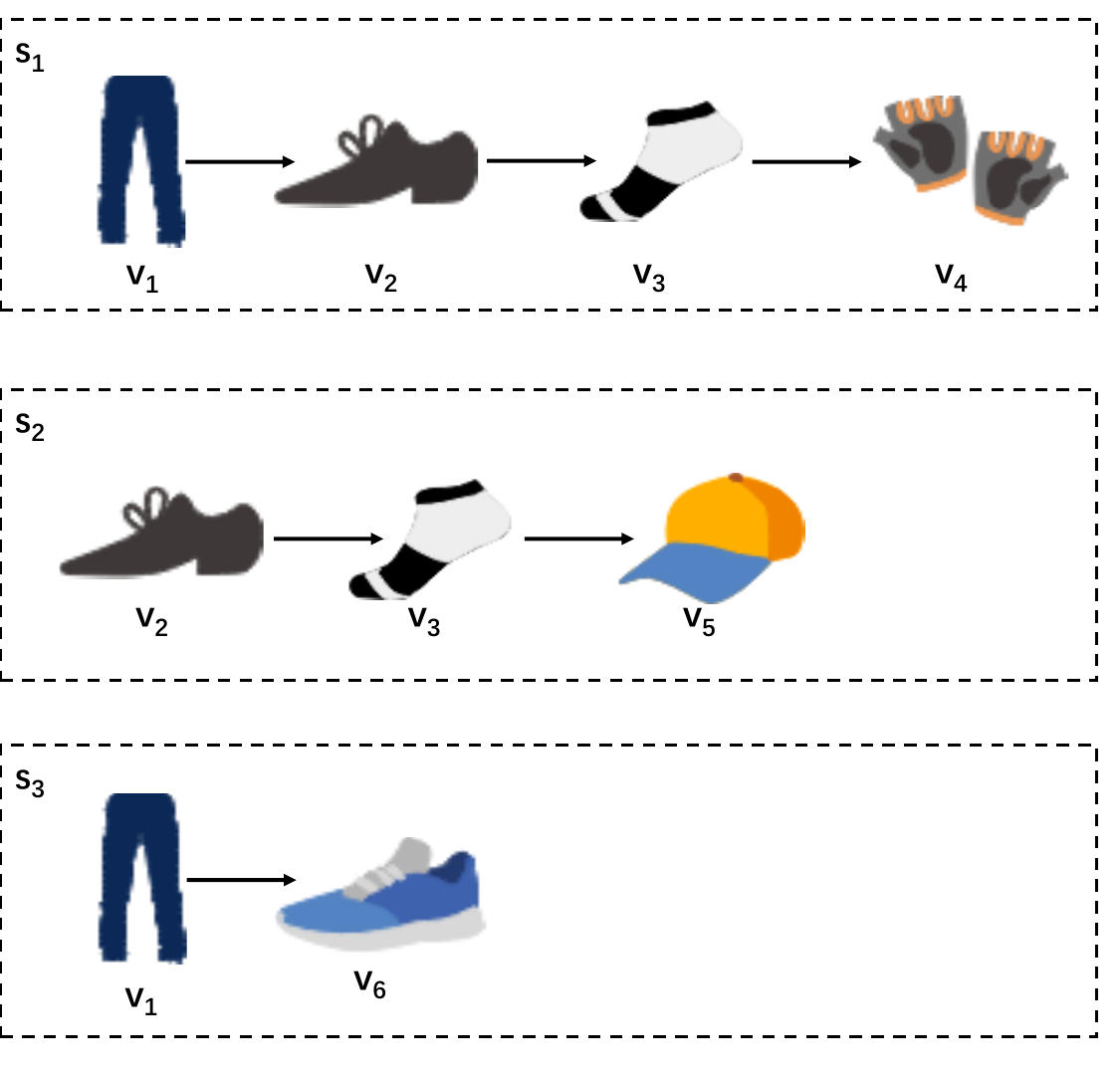}
\label{fig:1b}
}
\caption{Transitions of items and session sequence. (a)The transitions between items in the traditional recommendation system. (b)The session sequence in session -based recommendation system: the user behavior is divided into session sets. Here, the arrows indicate that users purchase the item v$_{i+1}$ after purchasing the item v$_i$.}
\label{fig:1}
\end{figure*}

In the era of big data, it is extremely difficult for users to select the information they need from a large number of products and services. The recommendation system can learn user preferences based on the user's historical data to help users make reasonable decisions and choices. With the development of recommendation systems and the analysis of user data, user preferences can be divided into long-term preferences and short-term preferences (\cite{wang2019a}). However, the traditional recommendation systems only considers the user's long-term preferences but ignores the transfer of user preferences. For example, in an e-commerce platform, the items purchased by users constitute the users' behavior sequence, and the purpose of the traditional recommendation system aims to learn users’ long-term preferences from transitions of items. Fig. \ref{fig:1} \subref{fig:1a} shows the transitions of six kinds of items. The traditional recommendation algorithm considers the transformational relations of items in all user behaviors so that the recommendation system can only focus on the items that the user pays attention to for a long time, which makes it impossible to quickly perceive the changes in users' interests. As a result, the recommendation system cannot quickly learn the transfer of user preferences. For example, a user who bought jeans for a long time suddenly likes sports pants. However, because the user has bought jeans more often than sports pants, the traditional recommendation system believes that the user's preference is still jeans. Therefore, the traditional recommendation system ignores the transfer of user preferences because it does not consider the transaction structure of user behavior. To solve this problem, it is necessary to decompose user behavior into smaller granularity, that is session. A session is a collection of an event or a group of events, which is transactional. For example, goods purchased by a user at one time, songs a user listens to in an hour, and web pages browsed within a day, can all be considered as a session (\cite{wang2019a}). The session is derived from the segmentation of user behavior into smaller granularity, which is endowed with transactional nature. The session-based recommendation method takes the session as the basic unit of recommendation, which can reduce the user's information loss, and it has been extensively studied.

As shown in Fig. \ref{fig:1} \subref{fig:1b}, in the e-commerce platform, the users’ purchasing behavior in Fig. \ref{fig:1} \subref{fig:1a} is divided into smaller granularity, and generate the session sequence \{s1, s2, s3\}. The session-based recommendation system decomposes user behavior into a set of sessions, thereby giving the user behavior transactional attribute (a session can be regarded as a transaction) so that the recommendation system can focus on the transfer of user’s preferences. For example, if a user has bought leather shoes for a long time, but recently prefers sports shoes, the session-based recommendation system can capture the transfer of user preferences promptly. Instead of continuing to recommend items related to leather shoes, as traditional recommendation systems do, it will recommend items related to sneakers. Thus, the session-based recommendation system ignores the long-term preferences of users. To make the session-based recommendation system not only focus on the short-term preferences of users but also consider the long-term preferences of users, it is necessary to fully consider the dependence between sessions. There are three main aspects of dependency in the conversation set: dependencies between different items in the same session, dependencies between different sessions, and dependencies between different items in different sessions. This requires that the session-based recommendation system model can fully consider the context of sessions and learn the complex transitions between items.

\begin{figure}[h]%
\centering
\includegraphics[scale=0.4,trim=60 0 80 0,clip ]{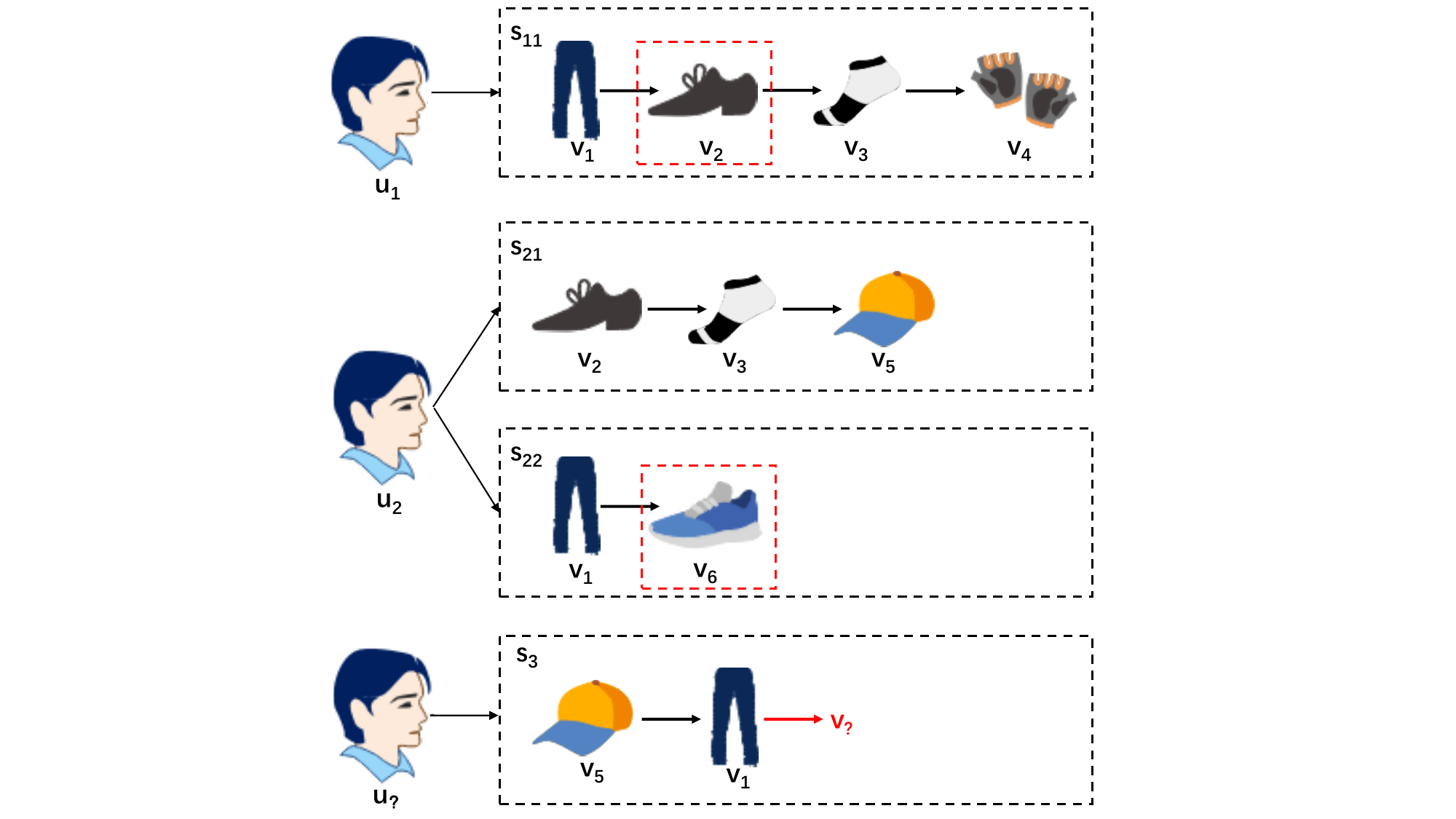}
\caption{Session sequence of different users. Here, the red wire frames show that different users may have different preferences after purchasing the item v$_1$.}
\label{fig:2}
\end{figure}

In recent years, graph neural networks have been applied to session-based recommendation systems. In the GNN based recommendation system, firstly, the session set is constructed as a directed graph, where the nodes of the directed graph represent items, and the edges represent the transfer of one item to another. Secondly, according to the constructed directed graph, the graph neural networks can learn the complex transformation relationship between items, and learn the item embeddings with strong expression ability, to generate session embeddings containing complex transformation information of items. For example, Session-based Recommendation with Graph Neural Networks, SR-GNN (\cite{wu2019session-based}), can not only capture the transfer of items in a short period but also consider the dependencies between distant items. Therefore, accurate item embedding vectors can be learned. Meanwhile, SR-GNN adds an attention network to focus on the user's local session embeddings and global session embeddings, so that the model can consider the user's long and short-term preferences. Although SR-GNN considers the long and short-term preferences of users and the complex transitions between items, it ignores other valid information in the session sequence, such as user information, which may lead to the loss of specific preferences of different users. For example, in Fig. \ref{fig:2}, two users have bought shoes after buying jeans, but they have different types of shoes. One user buys leather shoes and the other user buys sports shoes; u$_1$ buys gloves and u$_2$ buys hats. One new user (anonymous user) generates a session s$_3$, the two items in s$_{3}$ are related to u$_2$. When recommending the next item for the new user, u$_2$'s preference should be considered to a certain extent. It can be seen that certain user information can be captured in the item sequence in the session, which can improve the accuracy of the recommendation results.

This paper introduces HetG (\cite{sun2011pathsim}) to construct the user and other information that cannot be expressed in a homogeneous graph. HetG is a heterogeneous graph that contains multiple types of nodes and edges that represent different relationships. It can not only represent the transitions between nodes of the same type, but also the dependencies between different types of nodes, so the information that the heterogeneous graph can express is richer. Although HetG has a stronger ability to express data, it is very difficult to embed different nodes of the heterogeneous graph into a unified vector space. Heterogeneous Graph Neural Network, HetGNN (\cite{DBLP:conf/kdd/ZhangSHSC19}), learns latent vectors of different types of nodes in the heterogeneous graph according to the idea of aggregating heterogeneous neighbors. HetGNN can capture the heterogeneity of structure and content at the same time, and fully consider the transformation relationship of items.

To make the next item recommendation with rich and multi-type data information, we propose a novel method for Session-based Recommendation with Heterogeneous Graph Neural Networks, SR-HetGNN. SR-HetGNN takes full account of the dependencies among items, users, and sessions, and it can learn session embeddings with rich information and complex transitions of items.

The main contributions of this paper are as follows:

\begin{itemize}
\item We construct the session sequence into a heterogeneous graph, in which the rich dependency relationships among items, sessions, and users can be fully considered.
\item We propose a session-based recommendation with heterogeneous graph neural network, SR-HetGNN. SR-HetGNN 
can capture user's potential information from item sequence in anonymous session.

\item We conduct a large number of experiments on real-world datasets. The results shows that SR-HetGNN is superior to other existing methods.
\end{itemize}

\section{Related Work}\label{sec2}

In this section, we will first review the related work of a session-based recommendation system, including some traditional recommendation methods, and neural networks based recommendation methods. Then we will introduce the heterogeneous graph neural network.

\subsection{Traditional recommendation methods}
Collaborative filtering (\cite{barajas2005collaborative}) is one of the most popular recommendation methods. It categorizes users according to their ratings for items, and finds other users with similar interests for the target user. Then CF recommends items that other users are interested in but the target user has not seen or purchased to the target user. Matrix Factorization (MF) (\cite{DBLP:reference/sp/KorenB15}) is a typical traditional recommendation approach. It can embed users and items in the same vector space. The inner product of user embedding and item embedding is the user's degree of interest in the item, but MF cannot learn the sequence transformation of items. Therefore, Steffen Rendle \emph{et al.} combine Matrix Factorization and Markov Chain (MC), and realize the Factorized Personalized MC (FPMC) (\cite{rendle2010factorizing}). In this model, MF is used to learn user preferences, and MC uses transition graphs to model user sequence behavior so that FPMC can make the next recommendation better. Steffen rendele \emph{et al.} also propose a general optimization criterion BPR opt for personalized sorting, and apply it to MF algorithm to realize BPR-MF (\cite{rendle2009bpr:}).

\subsection{Neural-network-based methods}
In recent years, neural networks have been widely used in session-based recommendation methods. Hidasi B \emph{et al.} apply Recurrent Neural Network (RNN) to the recommendation system and propose a recommendation approach based on RNN, GRU4Rec (\cite{hidasi2015session-based}). GRU4Rec is essentially an improvement of the basic GRU model, it can consider the dependency between the previous node and the current node. GRU4Rec is sensitive to the user's sequence behavior, so that it can capture the transfer of user's preferences in time, but it cannot learn the user's long-term preferences. Li \emph{et al.} propose the Neural Attentive Recommendation Machine (NARM) (\cite{li2017neural}). NARM can learn users' short-term preferences and long-term preferences respectively with two RNNs. Liu \emph{et al.} propose Short-Term Attention/Memory Priority Model (STAMP) (\cite{liu2018stamp:}). This paper first proposes the short-term memory priority model (STMP). STMP can give priority to the user’s current interest preferences while considering the long-term preferences of users built outside the model. However, STMP may have the interest drift issue leading to incorrect recommendations. To address this issue, STMP adds an attention network to learn users' long-term preferences, that is the STAMP. STAMP can capture the user's long-term preferences in the long-term user behavior, and obtain the user's short-term preference through the latest click item in the current session, which achieves a good recommendation performance. Fang apply Self-Attention networks to session-based recommendation (SR-SAN), SR-SAN (\cite{DBLP:journals/corr/abs-2102-01922}) can capture the global dependencies among all items of a session regardless of their distance. Guo \emph{et al.} have implemented a Hierarchical Leaping Networks, HLN (\cite{DBLP:conf/sigir/GuoZFJP20}). HLN can explicitly establish the user's multiple preferences. It has a Leap Recurrent Unit (LRU), which is used to skip items that are not related to user preferences and accept learned preferences. HLN also has a Preference Manager (PM) to manage the learned preferences. HLN recommends items for users based on the users’ preferences in PM.

Graph Neural Network (GNN) (\cite{zhou2018graph,wu2019a,DBLP:journals/tois/QiuHLY20}) can capture the dependences between nodes through message passing between nodes,thus it can be applied to various fields, such as traffic planning(\cite{chen2021dynamic}). With the development of Graph Embedding (\cite{goyal2018graph}), Graph Neural Networks (GNNs) have also been widely used in recommendation systems. For example, Wang \emph{et al.} propose Global Context Enhanced Graph Neural Networks, GCE-GNN (\cite{DBLP:conf/sigir/Wang0CLMQ20}). GCE-GNN constructs local session graph and global session graph from local session sequence and global session sequence respectively and uses GNN to learn session-level item embedding and global item embedding. Here GNN can consider the complex transformation relationship between the items in the local session graph and the global session graph, which makes the learning item embeddings more effectively. GCE-GNN also combines session-level item embeddings and global item embeddings through the attention mechanism, and finally generates mixed session embeddings. Yu \emph{et al.} implement the Target Attentive Graph Neural Networks (TAGNN) (\cite{yu2020tagnn}). TAGNN constructs the session sequence into a session graph and learns the item embeddings with the GNN. TAGNN also designs a target-aware attention mechanism, which can capture the complex transformation relationship between items, and learn the changes of user's interest in different items at the same time. Star graph neural networks for session-based recommendation(SGNN-HN)(\cite{pan2020star}) uses star network to model the complex transition relationship between items in the ongoing session. In order to avoid over-fitting, highway network (HN) is used to adaptively select embedding from the project representation. Finally, aggregating the project embeddings in the ongoing session to represent the user's final preference for project prediction. SHARE(\cite{wang2021session}) based on hypergraph attention network, which uses limited evidence to accurately simulate user intentions in these short sessions.SGNN(\cite{WANG2022117114}) utilizes spatiotemporal data to accurately predict a user's next click based on their behavior patterns. The SGNN model is designed to effectively simulate the user's behavior over time and space, providing personalized recommendations in a conversational setting.Xia \emph{et al.} model session-based data as a hypergraph and then propose a dual channel hypergraph convolutional network named DHCN to improve session-based recommendation (\cite{DBLP:conf/aaai/0013YYWC021}). MIHSG(\cite{guo2022learning}) proposes to learn multi-granularity continuous user intention units, captures supplementary user intentions from session fragments, and reduces the burden of long dependency. In addition, the Intention Fusion Ranking (IFR) module is used to combine the recommendation results from user intentions of different granularity.

Compared with the traditional recommendation methods, the neural-network-based session recommendation methods are able to capture the complex dependencies among users, items and sessions. Therefore, more and more researchers apply different deep learning models to the session based recommendation and solve a series of problems,as shown in Table \ref{table2_1}.

\begin{table*}[htbp]\small
\centering
  \caption{Comparison of session recommendation models}
  \label{table2_1}
  \resizebox{\textwidth}{8mm}{
  \begin{tabular}{cl}
    \hline
    Model &  Problems Solved\\
    \hline
	 Traditional recommendation methods & Mining user preferences with display information \\
   	 Recurrent Neural Network &  Mining sequential patterns of user behavior using RNN  \\
     Attentional mechanism &  Emphasize the user's current purpose and ignore the user's wrong interactions       \\
     Graph Neural Network &   Consider complex structures and transformation relationship between items     \\
    \hline
    \end{tabular}}
\end{table*}

\subsection{Heterogeneous Graph Neural Network}
Heterogeneous Information Network (HIN) (\cite{DBLP:journals/tkdd/SunNHYYY13} \cite{DBLP:conf/asunam/YangWJX19}) can build multiple types of nodes into the same graph, and link these nodes with different types of edges, so that HIN can model complex context information (\cite{DBLP:journals/tkdd/SunNHYYY13}). However, it is difficult to embed these nodes or edges into the same vector space, because HIN has different types of nodes or edges. This paper calls HIN as Heterogeneous Graph (HetG) to compare with Homogeneous Graph.

Recently, graph representation learning (\cite{cui2019a}) has developed rapidly, and many graph embedding methods have been proposed, such as DeepWalk (\cite{perozzi2014deepwalk:}), metapath2vec (\cite{dong2017metapath2vec:}), node2vec (\cite{grover2016node2vec:}), and LINE (\cite{tang2015line:}), \emph{etc}. The DeepWalk algorithm generates node sequences by random walk, which is regarded as a sentence, and uses language modeling to generate the embedded representation of nodes. This paper uses the DeepWalk algorithm to embed heterogeneous graphs into the same vector space to obtain pre-embedding representations of various types of nodes. In recent years, HetGNN is also often applied to recommendation systems. Ren \emph{et al.} propose an effective citation recommendation method based on information network clustering, ClusCite (\cite{ren2014cluscite:}), which learns the relationship between citations in a heterogeneous information network and clusters them into interest groups. Hu \emph{et al.} develop a new deep neural network with a common attention mechanism: Meta-path based Context for Recommendation, MCRec (\cite{hu2018leveraging}). MCRec learns the effective representation of users, objects, and context-based on meta-paths, and it can realize powerful interactive functions. The ie-HGCN (\cite{yang2021interpretable})  is designed as a hierarchical aggregation architecture, consisting of object-level aggregation and type-level aggregation, to learn the representations of objects in HINs, demonstrating its superiority.HGNN-AC (\cite{jin2021heterogeneous}) employs the topological embedding of nodes as guidance and performs weighted aggregation of attributes from attributed nodes to complete the missing attributes for nodes without attributes. Extensive experimentation has demonstrated its superiority over baseline methods.
\section{Problem Definition}\label{sec3}
This section first defines the symbols that will be used, and then formally defines the studied problem.

\subsection{Symbol definition}
User behavior V=\{v$_{1}$, v$_{2}$, v$_{3}$, …, v$_{n}$\} is composed of the click sequence of all users U=\{u$_{1}$, u$_{2}$, u$_{3}$, …, u$_{h}$\}. Session-based recommendation system decomposes user behavior V into smaller granularity to obtain the session sequence S=\{s$_{1}$, s$_{2}$, s$_{3}$, …, s$_{m}$\}. A session s is composed of multiple items, s=\{v$_{s,1}$, v$_{s,2}$, v$_{s,3}$, …, v$_{s,i}$\}, where v$_{s,i}$ $\in$ V. To embed all nodes in the same model, it is necessary to embed all nodes into a vector space to obtain all node embeddings $Node=\{S^*, V^*, U^*\}$, where $S^*$ is session embedding set, $V^*$ is item embedding set, and $U^*$ is user embedding set. Common symbols can be seen in Table \ref{table3_1}.

\begin{table}[h]\small
\centering
  \caption{Symbol Table}
  \label{table3_1}
  \begin{tabular}{cl}
    \hline
    Notations& Descriptions\\
    \hline
    V&Item set contains all items\\
    S&Session set contains all sessions\\
    U&User set contains all users\\
 Arrt & Attribute set of a node\\
    v$_{i}$ &An item in item set\\
    s$_{i}$ &A session in session set\\
    u$_{i}$ &A user in user set\\
 arrt$_{i}$ &An attribute in attribute set\\
    $V^*$ & Item embedding set\\
    $S^*$ & Session embedding set\\
    $U^*$ & User embedding set\\
 $\bm{Arr}$ & Attribute embedding set\\
    $\bm{v_i}$ &An item embedding\\
    $\bm{s_i}$  &A session embedding\\
    $\bm{u_i}$ &A user embedding\\
    $\bm{arr_i}$ & An attribute embedding\\
    $Node$ &All node embedding set\\
    \hline
    \end{tabular}
\end{table}

\subsection{Problem definition}
\subsubsection{Heterogeneous graph Construction}
We construct the session sequence S into a heterogeneous graph G$_{n}$=(V, S, U, E$_{v}$, E). G$_{n}$ contains three types of nodes: item node V, session node S, and user node U, and two types of edges: the set of directed edges  between two items E$_{v}$, and the set of undirected edges E. E$_{v}$=(v$_{s,i}$, v$_{s,i+1}$) means that the user has purchased the next item v$_{s,i+1}$ after the user purchased item v$_{s,i}$. E$_{v}$ can represent the complex transformation relationship between items. E =\{(v$_{s,i}$, s), (v$_{s,i}$, u), (s, u)\}, represents the relationship between items and sessions, items and users, and users and sessions, respectively.

\subsubsection{Learning item embeddings with Heterogeneous Graph Neural Network}

After constructing the heterogeneous graph G$_{n}$, this paper designs the Heterogeneous Graph Neural Network model $F_\theta$ with parameters $\theta$. The model $F_\theta$ is used to learn the embedded representation of items, and the information of user nodes and session nodes are aggregated into the item embeddings $V^*=\{\bm{v_1}, \bm{v_2}, \bm{v_3}, …, \bm{v_i}\}$, where the item embedding $\bm{v_i}$ is the vector representation of node v$_{i}$.

\subsubsection{Session-based Recommendation with HetGNN}
The goal of session-based recommendation is to recommend the next item for the user in the current session. The model $F_\theta$ in this paper decomposes user behavior V into session sequence S and constructs session sequence S into heterogeneous graph G$_{n}$. Then the heterogeneous graph G$_{n}$ is embedded into the HetGNN model $F_\theta$, and it will learn item embeddings $V^*=\{\bm{v_1}, \bm{v_2}, \bm{v_3}, …, \bm{v_i}\}$, which contain rich information and transformational relations of items. Then the model will generate session embeddings $S^*=\{\bm{s_1}, \bm{s_2}, \bm{s_3}, …, \bm{s_m}\}$ according to the item embeddings $V^*$. Finally, the scores of items are obtained through the softmax layer, the top-n items will be recommended as the next item v$_{s,i+1}$ for the user.

\section{Methods}\label{sec4}

\begin{figure*}[htbp]
\centering
\includegraphics[scale=0.35]{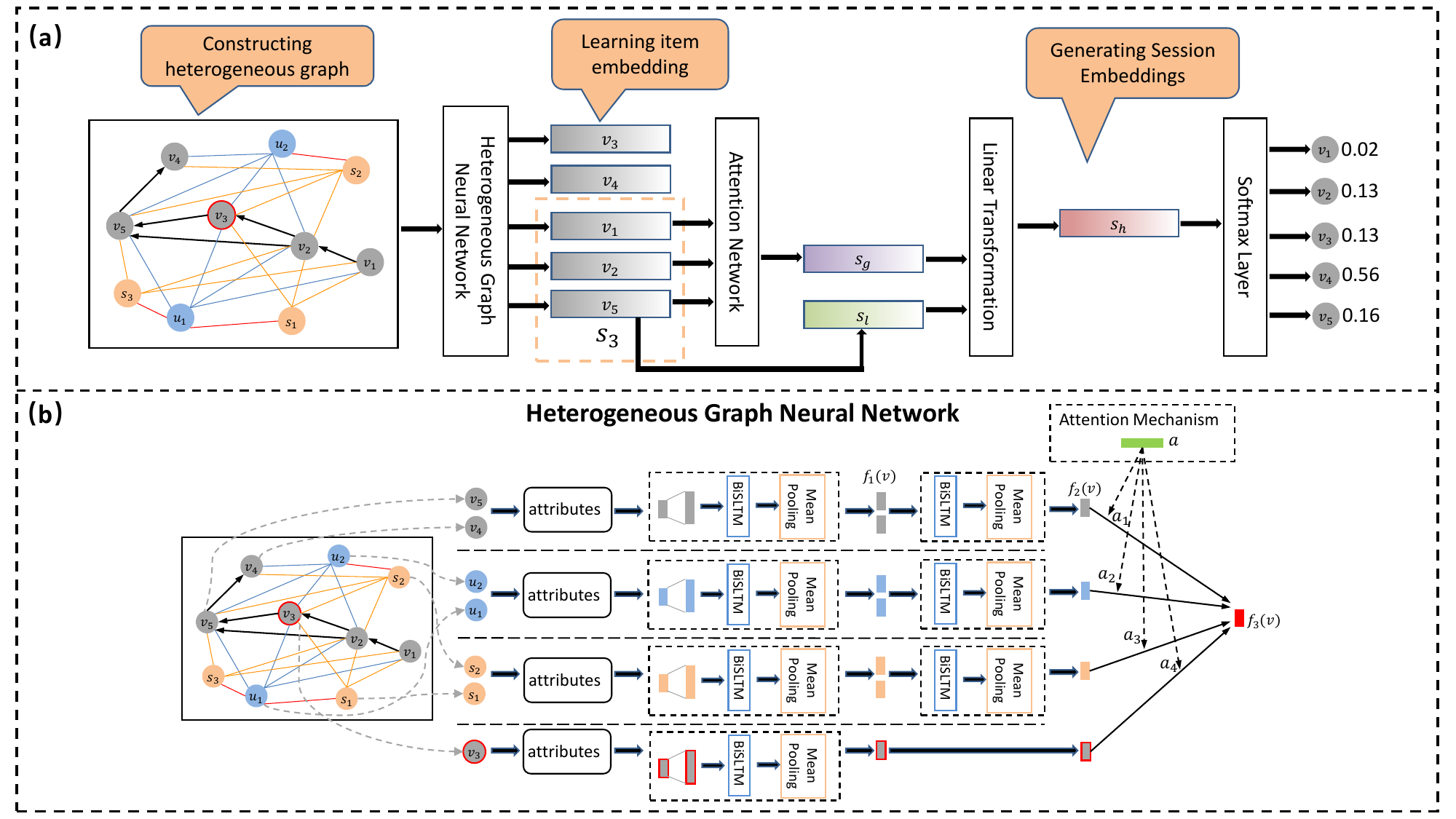}
\caption{Model structure diagram. (a)The model structure diagram of SR-HetGNN: Construct the session sequence into a heterogeneous graph, which contains three types of nodes: item nodes, session nodes, and user nodes; next, the item embeddings are learned through the HetGNN; then generate the session embeddings through the attention network, and finally give the recommendation result of the next item. (b) The structure of heterogeneous graph neural network(HetGNN): first, sample heterogeneous neighbors for all item nodes; next, aggregate the content of heterogeneous neighbor nodes; then aggregate the heterogeneous neighbors of the same type; finally aggregate different types to obtain the final embedding of item nodes.}
\label{fig:3}
\end{figure*}

This section will introduce the model structure of SR-HetGNN, SR-HetGNN is mainly divided into three parts:  constructing the session sequence into a heterogeneous graph; learning the item embeddings; and generating the session embeddings. Fig. \ref{fig:3} shows the model structure of SR-HetGNN.

\subsection{Building the Heterogeneous Graph}
First, session sequence S needs to be constructed into a Heterogeneous Graph G$_{n}$=(V, S, U, E$_{v}$, E), G$_{n}$ is shown in Fig. \ref{fig:3} (a), and its structure is described in Section 3.2. After G$_{n}$ is created, the DeepWalk algorithm (\cite{perozzi2014deepwalk:}) is used to embed all nodes in the heterogeneous graph into a vector space. The steps of the DeepWalk algorithm are as follows: First, the random walk with fixed step length $l$ is carried out from each node to get the word vectors $Word=\{word_1, word_2, word_3, …, word_n\}$, where $word_i$ is the symbolic representation of a node in G$_{n}$. Then Word2vec is used to train these word vectors to generate the pre-embedding vectors $Node=\{S^*, V^*, U^*\}$ for all nodes. For example, as shown in Fig. \ref{fig:3} (a), the random walk of the DeepWalk starts at node v$_{2}$, and walk to node u$_{2}$ randomly, then jumps to node v$_{4}$, and repeat the random walk until the word vector $Word$=\{v$_{2}$, u$_{2}$, v$_{4}$, …\} with length $l$ is obtained. Finally, Word2vec trains the word vectors to get the pre-embedding vectors of all nodes.

\subsection{Learning item embeddings}
After the pre-embedding node vectors are generated, each node in the heterogeneous graph G$_{n}$ can be represented by vectors of the same dimension. However, the pre-embedding node vectors lack the expressive ability and cannot express the complex transitions between items. Therefore, this paper uses Heterogeneous Graph Neural Network, HetGNN, to learn item embeddings that contain rich information and complex transformation relationships. The model structure of HetGNN can be seen in Fig. \ref{fig:3} (b). The core idea of HetGNN is aggregation, which is mainly divided into four steps: Sampling heterogeneous neighbors, aggregating the content of the node's heterogeneous neighbors,  aggregating heterogeneous neighbors of the same type, and aggregating different types.

\subsubsection{Sampling heterogeneous neighbors}
An important issue for learning item embeddings is how to aggregate heterogeneous neighbor nodes with different content. The types and numbers of these heterogeneous neighbor nodes are different, and aggregating these nodes may require different feature transformations. For example, a user purchases three items in two sessions, and another user purchases four items in three sessions. Here, the sizes of neighbor nodes of the two users are different. It is an important problem to select nodes as heterogeneous neighbors of users so that they can use the same model to aggregate these heterogeneous neighbors. To solve this problem, this paper adopts the restart-based random walk (RWR) (\cite{wu2019session-based}) method to sample heterogeneous neighbors. The main steps of RWR are as follows:

\begin{enumerate}[Step 1: ]
  \item For all item nodes V=\{v$_{1}$, v$_{2}$, v$_{3}$, …, v$_{n}$\}, start random walkings from each item node v$_{i}$. In the process of random walk, there is a probability $P$ that it will return to the initial node v$_{i}$. In order to get all types of nodes during the random walk, RWR controls the number of each type of nodes to walk, and stores all the walked nodes in a list, namely RWR(v);
  \item Classify all nodes in RWR(v), and select top-k$_{t}$ nodes for each type t as heterogeneous neighbors of item node v$_{i}$. This method can select all types and the same number of heterogeneous neighbors for each item node.
\end{enumerate}

\subsubsection{Aggregating the node content of heterogeneous neighbors}
Different types of heterogeneous neighbors have different content of nodes. For example, user nodes may contain attributes such as age and gender, and item nodes contain attributes such as item name and type. Thus how to encode the different attributes of nodes into fixed-dimensional embedded representations through neural networks? HetGNN designs an architecture based on Bi-directional LSTM (BiLSTM) to obtain the interaction between features and aggregate all the attributes of the node into the embedded representation of the node so that it has greater representation ability. The specific steps are as follows:

First, node v is a heterogeneous neighbor of node v$_{i}$, its attribute set is Arrt=\{arrt$_{1}$, arrt$_{2}$, …, arrt$_{n}$\}. We use different models to transform the attribute arrt$_{i}$ into the embedding vector of the same dimension $\bm{Arr}=\{\bm{arr_1}, \bm{arr_2}, …, \bm{arr_n}\}$, HetGNN provides different solutions for different types of attributes, such as using one-hot for text attributes and CNN for images. After getting the embedding vector of each attribute of the node, the node embedding $\bm{f_1 (v)}$ can be formulated as follows:

\begin{equation}
\bm{f_1 (v)}=\frac{\sum\limits_{\bm{arr} \in \bm{Arr}}\{\overrightarrow{LSTM}[FC(\bm{arr})]\oplus \overleftarrow{LSTM}[FC(\bm{arr})]\}}{\|\bm{Arr}\|}
\label{1}
\end{equation}

\noindent where $\bm{f_1 (v)} \in R^{d\times1}$, $d$ is the dimension of node embedding; $FC$ is the vector transformation layer, which can be a fully connected layer; $\oplus$ is the connection operation. The formula of BiLSTM is as follows:

\begin{equation}
\begin{split}
   &\bm{z_i}=\sigma (\bm{U_z} FC_\theta (\bm{x_i})+ \bm{W_z h_{i-1}} + b_z) \\
   &\bm{f_i}=\sigma (\bm{U_f} FC_\theta (\bm{x_i})+ \bm{W_f h_{i-1}} + b_f) \\
   &\bm{o_i}=\sigma (\bm{U_o} FC_\theta (\bm{x_i})+ \bm{W_o h_{i-1}} + b_o) \\
   &\bm{\hat{c}_i}=\tanh (\bm{U_c} FC_\theta (\bm{x_i})+ \bm{W_c h_{i-1}} + b_c) \\
   &\bm{c_i}= \bm{f_i} \circ \bm{c_{i-1}}+\bm{z_i} \circ \bm{\hat{c}_i} \\
   &\bm{h_i}=\tanh (\bm{c_i})\circ \bm{o_i}
\end{split}
\label{2}
\end{equation}

\noindent where $\bm{U_z}, \bm{W_z}$, \emph{etc.} are learning parameters, $\bm{z_i}, \bm{f_i}$, and $\bm{o_i}$ are the forget gate vector, input gate vector and output gate vector; $\bm{h_i}$ is the output hidden state.

This method can aggregate heterogeneous content to make node embeddings more expressive, and it is convenient to add content attributes of nodes.

\subsubsection{Aggregating heterogeneous neighbors of the same type}
After aggregating the node contents of heterogeneous neighbors, their embedding representations of heterogeneous neighbors are generated. Each node has multiple types of heterogeneous neighbors, and each type $t$ has multiple heterogeneous neighbors. HetGNN has designed a neural network to aggregate the same type of nodes into an embedding vector. This part still uses BiLSTM to aggregate nodes of the same type and learn the complex relationships between them, so that the learned type embedding $\bm{f_2 (t)}$ has a stronger expressive ability. The calculation of type embedding $\bm{f_2 (t)}$ is formulated as follows:

\begin{equation}
\bm{f_2 (t)}=\frac{\sum\limits_{\bm{v} \in t}\{\overrightarrow{LSTM}[\bm{f_1(v)}]\oplus \overleftarrow{LSTM}[\bm{f_2(v)}]\}}{|t|}
\label{3}
\end{equation}

\noindent where the BiLSTM model is the same as Formula (\ref{2}).

\subsubsection{Aggregating different types}
After getting the type embedding, we need to aggregate all types of embedding into a vector, which is the final embedding of node v$_{i}$. However, different types of heterogeneous neighbors have different effects on node v$_{i}$, so the attention mechanism is introduced. The calculation formula for the importance of different types of node v$_{i}$ is as follows:

\begin{equation}
a^{v_i,t}=\frac{exp\{ LeakyReLU(\bm{U}[\bm{f_2} (t)\oplus \bm{f_1 (v_i )}]\}}{\sum\limits_{\bm{f_j}\in F_2 (T)\bigcup \bm{f_1 (v_i)}}exp\{ LeakyReLU(\bm{U}[\bm{f_j}\oplus \bm{f_1 (v_i )}]\}}
\label{4}
\end{equation}

\noindent where $LeakyReLU$ is the leaky version of a Rectified Linear Unit, and $\bm{U}\in R^{1\times2d}$ is the attention parameter. $T$ is the typeset of heterogeneous nodes, $F_2(T)$ is the set of $\bm{f_2(t)}$, and $t$ is a heterogeneous node type in $T$. After calculating the importance of each type to node v$_i$, the final embedding of node v$_i$ is calculated as follows:

\begin{equation}
\bm{f_3 (v_i)}=\sum\limits_{\bm{f_j}\in \bm{f_2} (T)\bigcup \bm{f_1} (\bm{v_i})}a^{v_{i,j}} \bm{f_j}
\label{5}
\end{equation}

As the final embedding of the item node $\bm{v_i}=\bm{f_3}(\bm{v_i})$, it not only contains the transitions between items but also learns the information of other types of nodes to make it more expressive.

\subsection{Generating Session Embeddings}
After learning the item embeddings with HetGNN, the session embeddings can be generated. For session s=\{v$_{s,1}$, v$_{s,2}$, v$_{s,3}$, …, v$_{s,i}$\}, the calculation of pre-embedding vector is shown as follows:

\begin{equation}
\bm{s}=\bm{v_{s,1}}\oplus \bm{v_{s,2}}\oplus … \oplus \bm{v_{s,i}}\oplus \bm{v_0}
\label{6}
\end{equation}

\noindent where $\bm{v_{s,i}}$ is the node embedding of node v$_{s,i}$; $\bm{v_0}$ is zero vector without a fixed dimension; and $\oplus$ is connection operation. Since the number of items in a session may be different, $\bm{v_0}$ needs to be connected so that all pre-embedding session vectors have the same dimension.

Since the user’s long-term preferences and short-term preferences have different effects on the recommendation results, the attention mechanism is added to the model to obtain a hybrid session embedding that can express long-term and short-term preferences. In this paper, we first consider the local embedding $\bm{s_l}$ of this session. $\bm{s_l}$ is formulated as follows:

\begin{equation}
\bm{s_l}=\bm{v_{s,n}}
\label{7}
\end{equation}

\noindent where $\bm{v_{s,n}}$ is the embedding vector of the last item in the current session.

For the user's long-term preferences, it is necessary to consider the transformation relationship between all items. In this paper, a soft attention mechanism is used to learn the global embedding $\bm{s_g}$. The global session embedding $\bm{s_g}$ is formulated as follows:

\begin{equation}
\begin{split}
     & a_i=\bm{W^{T}}\sigma( \bm{W_1 v_{s,n}} + \bm{W_2 v_{s,i}}+c) \\
     & \bm{s_g}=\sum_{i=1}^{n}a_i \bm{v_{s,i}}
\end{split}
\label{8}
\end{equation}

\noindent where the matrices $\bm{W^T}\in R^d, \bm{W_1, W_2}\in R^{d\times2d}$ are the weights used to generate item embeddings. After obtaining the global embedding and local embedding of the session, the hybrid session embedding can be formulated as follows:

\begin{equation}
\bm{s_h}=\bm{W_3}[\bm{s_l;s_g}]
\label{9}
\end{equation}

\noindent where the matrix $\bm{W_3} \in R^{d\times2d}$ is used to fuse $\bm{s_l}$ and $\bm{s_g}$ into the hybrid session embedding.

After generating session embedding, the score of each candidate item is calculated through the softmax layer, and then the model is trained by a backpropagation algorithm. The detailed training process of the model is shown in Algorithm \ref{algorithm1}.

\renewcommand{\algorithmicrequire}{\textbf{Input:}}  
\renewcommand{\algorithmicensure}{\textbf{Output:}} 

\begin{algorithm}[H]	
\setstretch{1.35}
 \caption{The training process with SR-HetGNN}\label{algorithm1}
  \begin{algorithmic}
    \Require
      Session set S, User set U, Item set; Number of user neighbors, Number of item neighbors,  Number of session neighbors.
     \State Initialize the model of SR-HetGNN $F_{\theta}$.
       \State Build the Heterogeneous Graph G$_{n}$=(V, S, U, E$_{v}$, E).
       \State Generate the pre-embedding vectors $Node=\{S^*, V^*, U^*\}$ for all nodes by DeepWalk.
        \For {$t = 1, 2, 3, ..., T$}  
        \State Sample a mini-batch of session sequence S$_h$ = \{s$_1$, s$_2$, s$_3$,..., s$_h$ \}.
        	\For {each s$_h \in$ S}
       		   \For{each v$_i \in$ s$_h$}  
		 	    \State Sample heterogeneous neighbors by Restart-based Random Ralk (RWR).
			    \State Get the aggregated embeddings of each  heterogeneous node by Formula (\ref{1})
			    \State Get the aggregated embeddings of each  heterogeneous type by Formula (\ref{3})
			    \State Calculate attention factor $a^{v_i,t}$ of each  heterogeneous type by Formula (\ref{4}).
			    \State Get  final embedding of item $\bm{v_i}$ by Formula (\ref{5}).
		    \EndFor  
	       \State  Get local session embedding $\bm{s_l}$ by Formula (\ref{7}).
	       \State Get global session embedding $\bm{s_g}$ by Formula (\ref{8}).
	       \State Generate final session embedding  $\bm{s_h}$ by Formula (\ref{9}).
 	     \EndFor  
	    \State  Calculate scores candidate items through the Softmax Layer.
	    \State  Calculate $Loss$ by cross entropy loss function.
	    \State  Updata $\theta$ of the model of SR-HetGNN $F_{\theta}$.   
        \EndFor
  \end{algorithmic}  
 \label{algorithm1}
\end{algorithm}

\section{Experiments and results}\label{sec5}
In this section, we first introduce the data set used in the experiment, baselines for comparison, and evaluation metrics. Then, we give the experimental results and analyze the results.
\subsection{Dataset}
Two real-world datasets, Diginetica$\footnote{http://cikm2016.cs.iupui.edu/cikm-cup}$ and Tmall$\footnote{https://tianchi.aliyun.com/dataset/dataDetail?dataId=42}$, are used to evaluate the performance of the proposed model.

Diginetica comes from CIKM Cup 2016. Firstly, Diginetica is preprocessed to remove the data of anonymous users based on its transaction data. Because user nodes are needed when constructing heterogeneous graphs, anonymous users’ records in the data should be deleted. Secondly, we delete items that appear less than 5 times in the data set, and all sessions with only one item in the session. Finally, we split the data set and use the last few days of data as the test set, and the others as the training set. At the same time, users who do not exist in the train set are deleted in the test set to verify the influence of user information on the recommendation results, and the items that do not exist in the train set are also deleted. The final data set is as Table \ref{table5_1}.

\begin{table}[h]
\centering
  \caption{Dataset}
  \label{table5_1}
  \begin{tabular}{lll}
    \hline
    Statistics& Diginetica& Tmall\\
    \hline
    \# of items&16,882&10,513\\
    \# of training sessions&149,295&75,348\\
    \# of testing sessions&2,346&9,175\\
    \# of users&42,205&13,076\\
    \hline
    \end{tabular}
\end{table}

Tmall dataset comes from the Tianchi Dataset of Alibaba Cloud. This paper selects the add-to-favorite data of users in Tmall dataset within 4 months. Firstly, the Tmall dataset is preprocessed to remove the users whose operations are less than 20, the items that appear less than 10 times, and the session whose length is 1. Then, the dataset is divided and the data of the last 15 days is selected as the test set. As with the Diginetica data, users and items that do not exist in the train set are removed from the test set. The final dataset after preprocessing resulting dataset is shown in Table \ref{table5_1}.

\subsection{Baselines}
To evaluate the performance of this model, we compare our model with the following baselines:

\begin{itemize}
\item BPR-MF (\cite{rendle2009bpr:}) : It uses Matrix Factorization(MF) to learn user preferences, and it proposes a general optimization criterion BPR-opt for personalized sorting and applies it to MF.
\item FPMC (\cite{rendle2010factorizing}) : Matrix Factorization (MF) is used to learn users' preferences, and Markov Chain (MC) is used to model user's sequential behavior by transition graph.
\item GRU4Rec (\cite{hidasi2015session-based}) : It is an RNN-based recommendation model, an improvement of the basic GRU, and it is sensitive to the item sequence.
\item SR-GNN (\cite{wu2019session-based}) : It constructs the session sequence into a homogeneous graph and uses Graph Neural Network(GNN) to learn item embeddings.
\item TA-GNN (\cite{yu2020tagnn}) : It corporates the target-aware attention mechanism into GNN model to consider user interests given a certain target item as well as complex item transitions in session.
\item SHARE (\cite{wang2021session}) : It applies hypergraph attention network to accurately simulate user intentions in short sessions.
\item GCE-GNN(\cite{wang2020global}):  GCE-GNN is designed to create a local session graph and a global session graph from a given local session sequence and global session sequence, respectively.
\item MIHSG (\cite{guo2022learning}) : Using multi-granularity continuous user intention units to captures supplementary user intentions from session fragments.
\end{itemize}

\begin{figure}[htbp]
\centering
\subfloat[User type]{
\includegraphics[scale=0.55, trim=330 155 330 155, clip ]{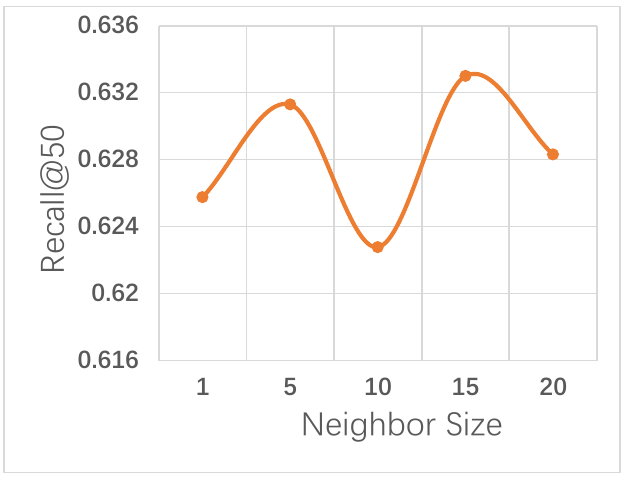}
\label{fig:41}
}
\subfloat[Item type]{
\includegraphics[scale=0.55, trim=330 155 330 155, clip ]{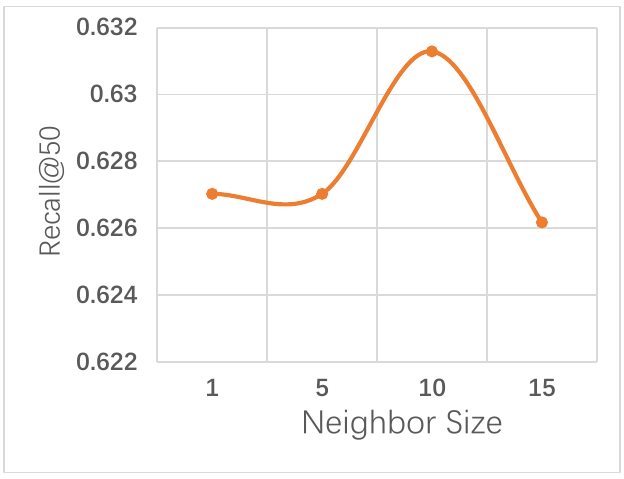}
\label{fig:42}
}\\
\subfloat[Session type]{
\includegraphics[scale=0.55, trim=330 155 330 155,clip ]{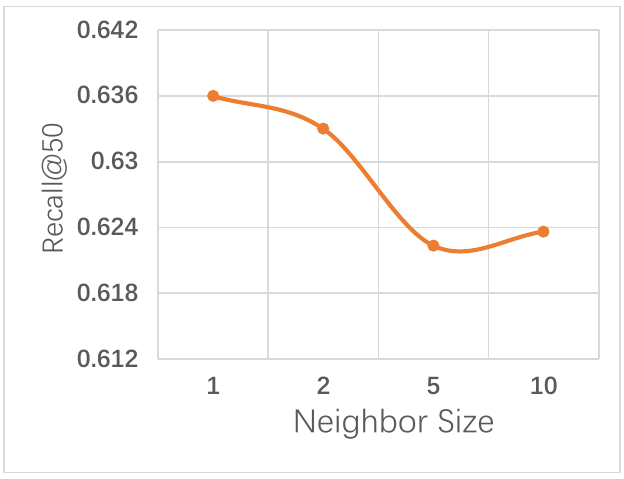}
\label{fig:43}
}
\caption{Performance w.r.t. sampled neighbor size on Diginetica.}
\label{fig:4}
\end{figure}
\subsection{Evaluation Metrics}
This paper uses the following evaluation metric to evaluate the model.

$Recall@50$: Recall is widely used as a measure of recommendation. It represents the percentage of correctly recommended items in the sample.

$MRR@50$: MRR (Mean Reciprocal Rank) is the average of reciprocal ranks of the correctly recommended items, and the reciprocal rank is set to 0 when the rank exceeds n. MRR considers the order of recommendation ranking.

\begin{figure}[htbp]
\centering
\subfloat[User type]{
\includegraphics[scale=0.55,trim=330 155 330 155,clip ]{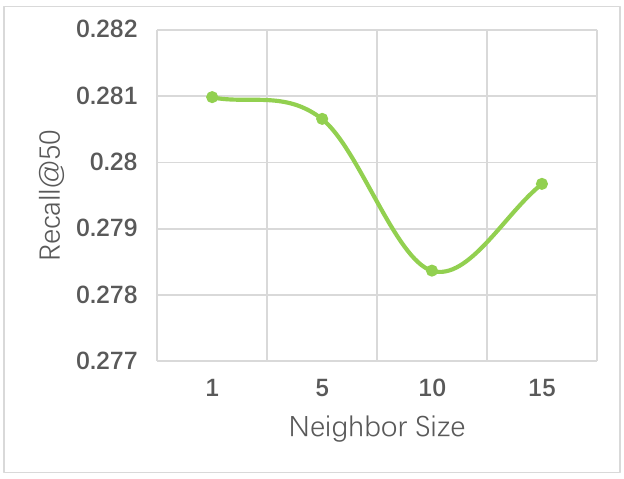}
\label{fig:51}
}
\subfloat[Item type]{
\includegraphics[scale=0.55,trim=330 155 330 155,clip ]{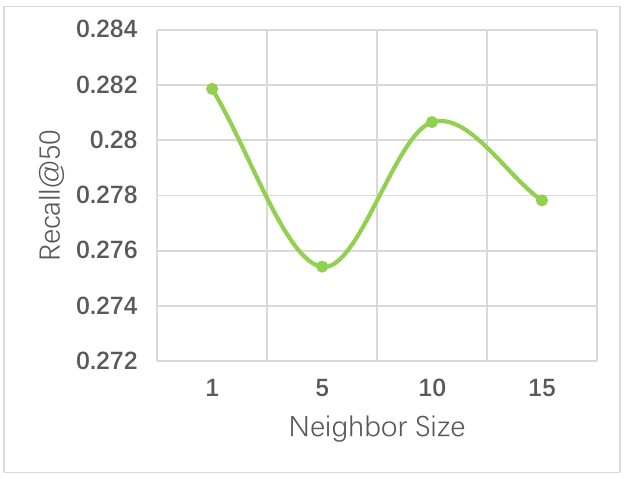}
\label{fig:52}
}\\
\subfloat[Session type]{
\includegraphics[scale=0.55,trim=330 155 330 155,clip ]{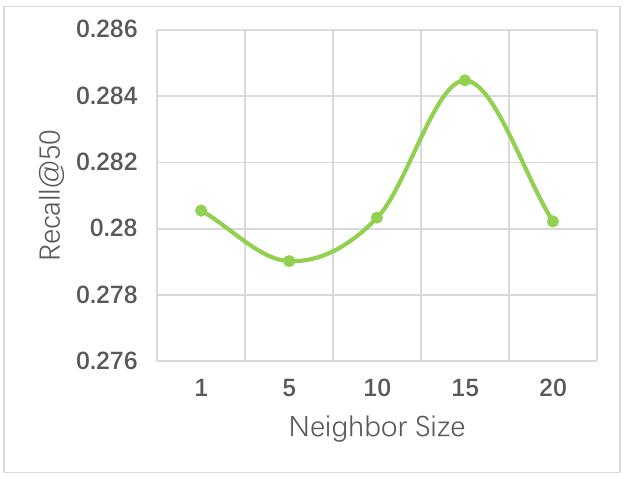}
\label{fig:53}
}
\caption{Performance w.r.t. sampled neighbor size on Tmall. }
\label{fig:5}
\end{figure}

\subsection{Parameter setup}
Hyper-parameters are very important for model training. To a certain extent, they determine the quality of model training results. In our model, the number of heterogeneous neighbors of an item node is a very important parameter. This set of parameters includes the number of user neighbors, the number of item neighbors, and the number of session neighbors. This paper selects the optimal values of this set of parameters through a large number of experiments on Diginetica and Tmall dataset, and draws the curves based on the experimental results, as shown in Fig. \ref{fig:4} and Fig. \ref{fig:5}.

According to Fig. \ref{fig:4}, on the Diginetica dataset, the performance of the model is better when the number of user neighbors is 15, the number of item neighbors is 10, and the number of session neighbors is 1. However, it is different from Diginetica dataset that the model performs better on Tmall dataset when the number of user neighbors is 1, the number of item neighbors is 1, and the number of session neighbors is 15. The possible reason is that the model needs a different number of neighbors to extract information in different length session sequences. For example, on Diginetica dataset, the optimal value of the number of user neighbors is 15, while the optimal value of the number of user neighbors is 1 on Tmall dataset. It can be seen that, in different scenarios, the model uses too many heterogeneous neighbors to extract information, which may be disturbed by noisy nodes, thus reducing the performance of the model.

\begin{figure}[h]
\centering
\subfloat[Performance on Diginetica]{
\includegraphics[scale=0.31,trim=220 110 220 110,clip ]{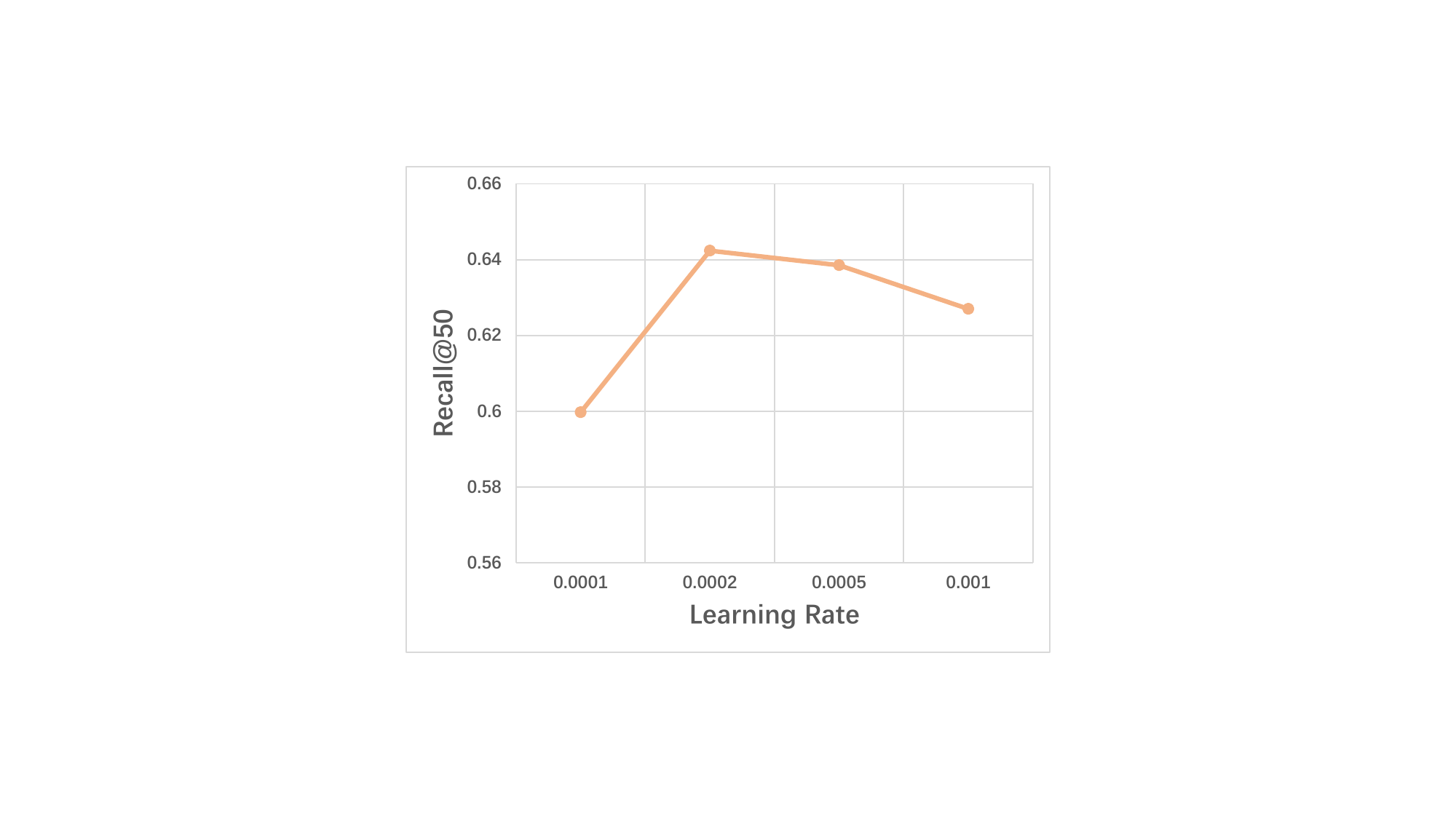}
\label{fig:lra}
}
\subfloat[Performance on Tmall]{
\includegraphics[scale=0.31,trim=220 110 220 110,clip ]{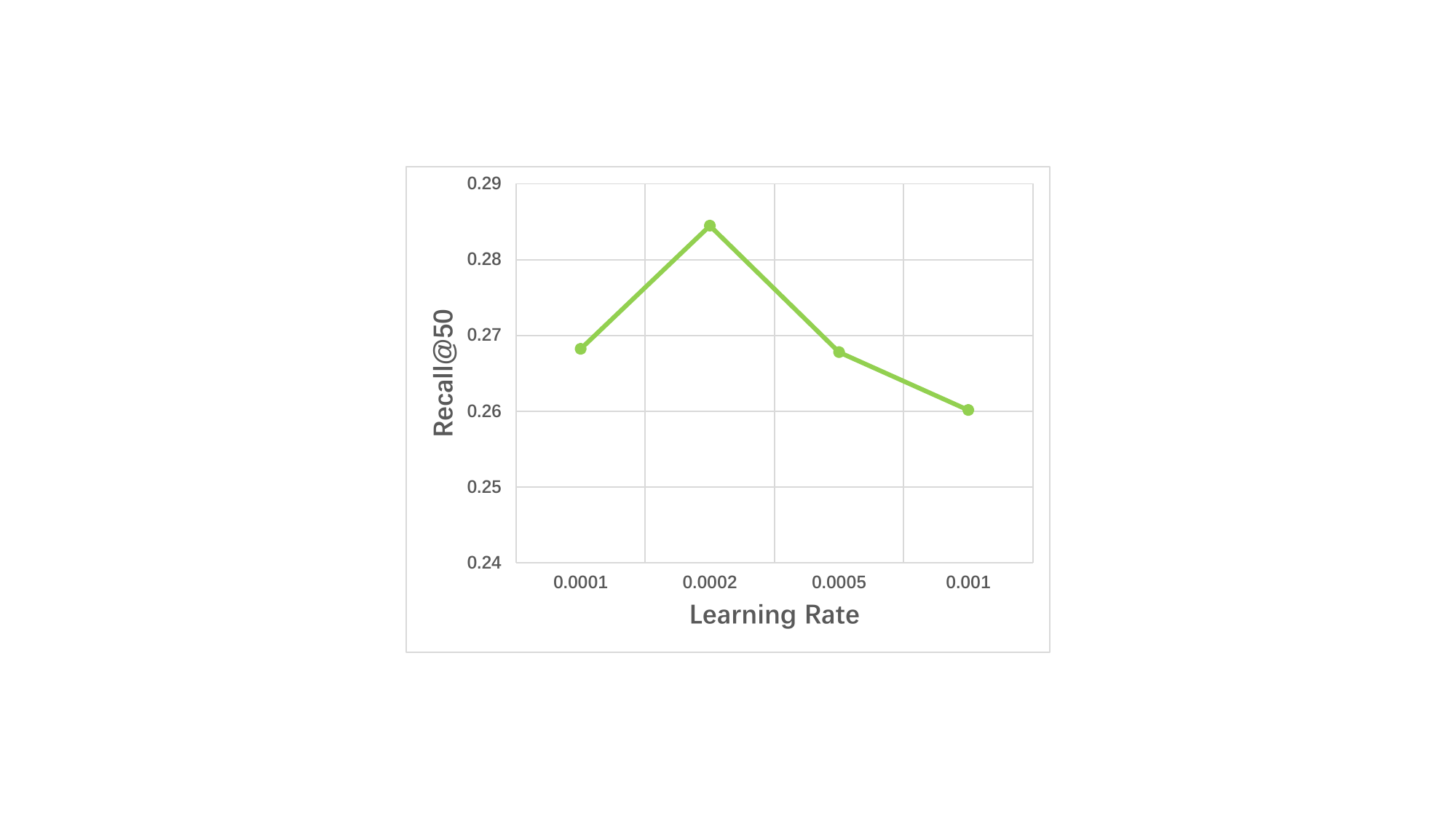}
\label{fig:lra}
}
\caption{Performance w.r.t. learning rate.}
\label{fig:6}
\end{figure}

As an important hyper-parameter in deep learning, the learning rate (lr) determines whether the objective function converges to the local minimum and when it converges to the minimum. In this paper, different learning rates are set to train the model, the final results as shown in Fig. \ref{fig:6}. It can be seen from the figure that when lr$=0.0002$, the performance of SR-HetGNN is best on both Diginetica and Tmall datasets. Therefore, lr $= 0.0002$ is chosen as the optimal parameter of the learning rate.

\begin{figure}[h]
\centering
\subfloat[Comparison on Diginetica]{
\includegraphics[scale=0.31,trim=220 110 220 110,clip ]{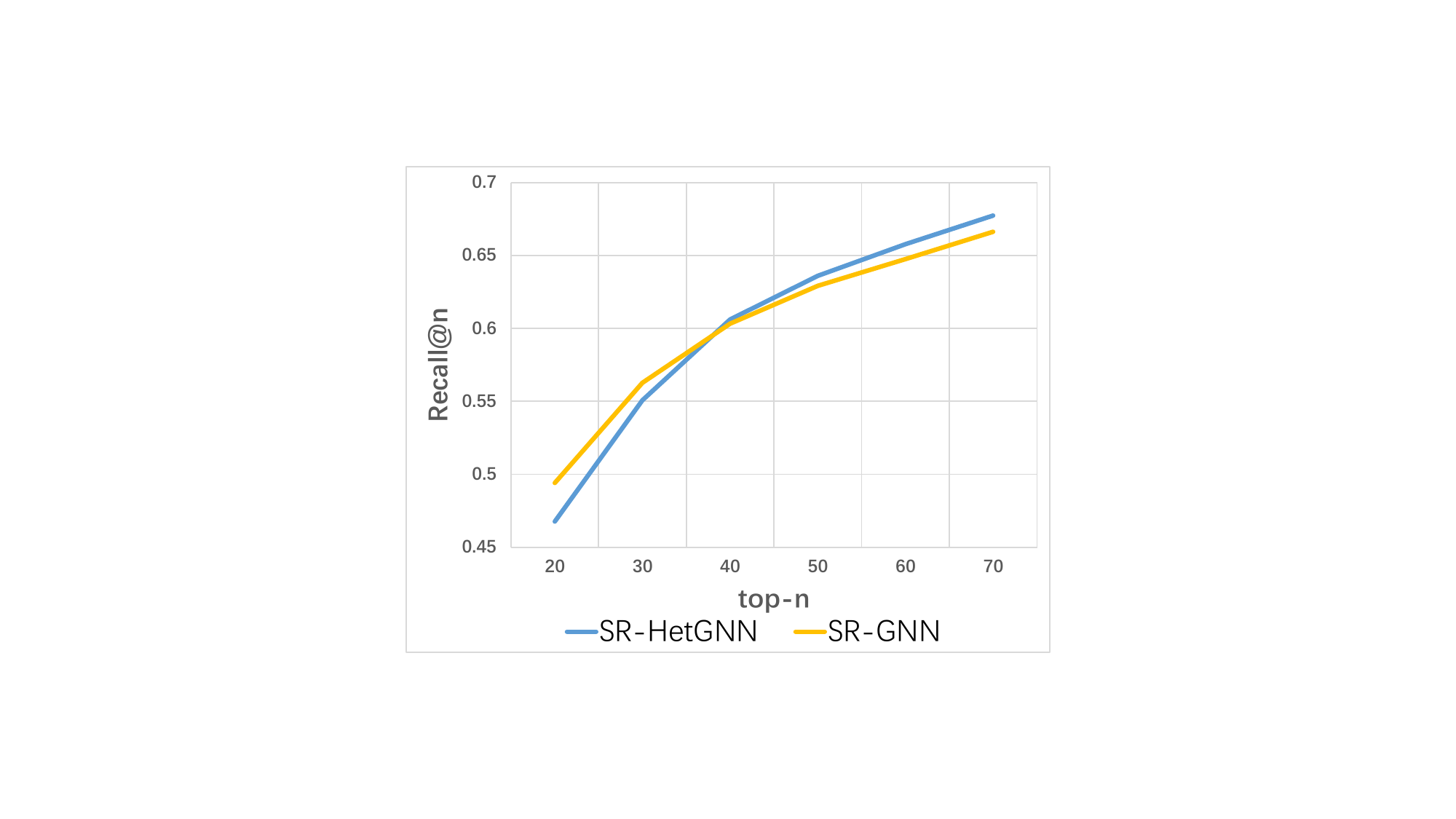}
\label{fig:topna}
}
\subfloat[Comparison on Tmall]{
\includegraphics[scale=0.31,trim=220 110 220 110,clip ]{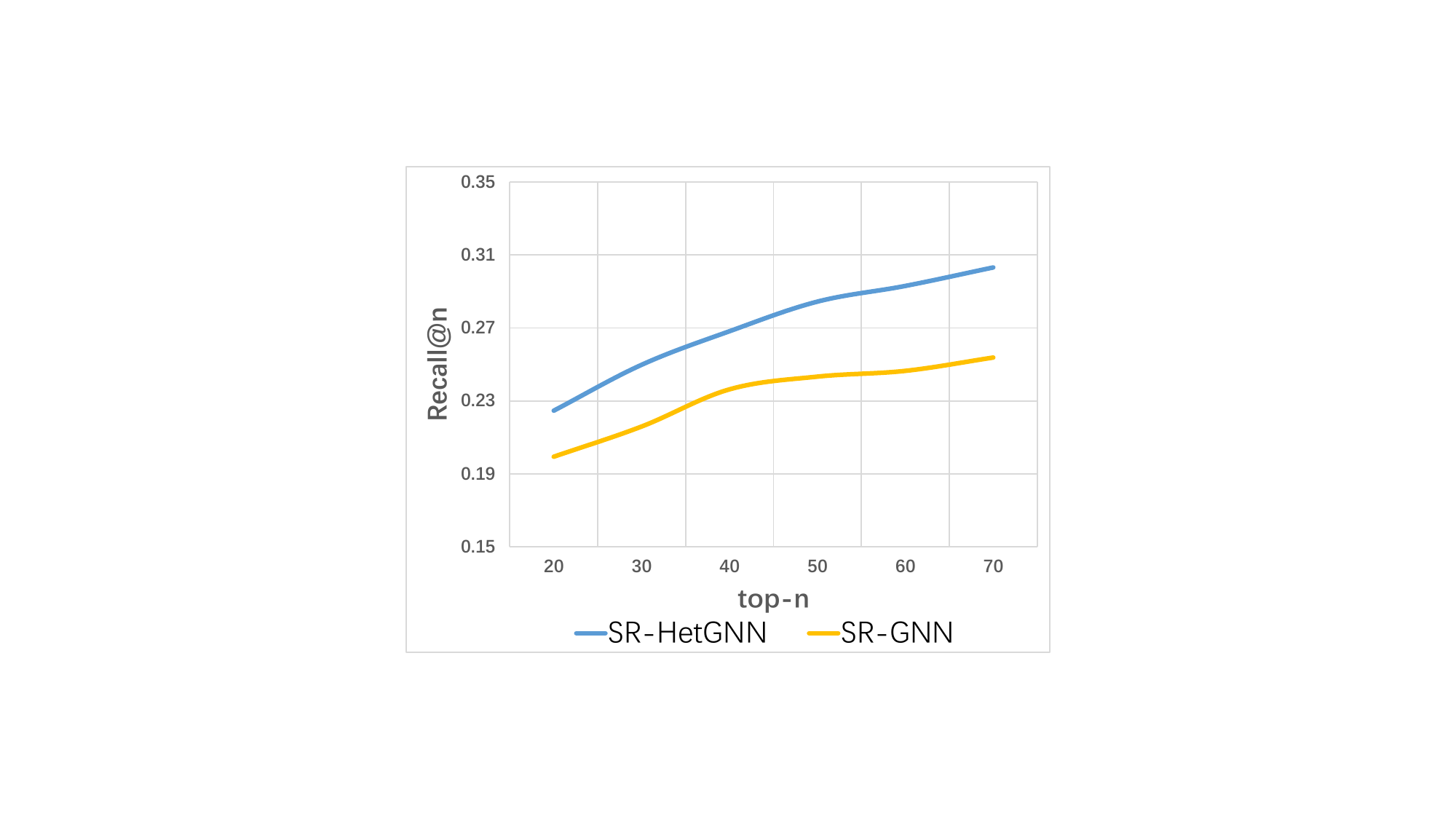}
\label{fig:topnb}
}
\caption{Comparison of top-n recommendation results.}
\label{fig:7}
\end{figure}

To understand the effect of $n$ value in top-n on the recommendation performance, this paper compares with SR-GNN in different top-n recommendations, and draws the curve according to the experimental results in Fig. \ref{fig:7}. It can be seen from Fig. \ref{fig:7} \subref{fig:topna} that Recall@n of SR-GNN is higher than that of SR-HetGNN when n$<$40, Recall@n of SR-HetGNN is better than that of SR-GNN when n$\geq$40 on Diginetica. Fig. \ref{fig:7} \subref{fig:topnb} shows that the performance of SR-HetGNN is always superior to that of SR-GNN on Tmall. With the increase of $n$, the performance advantage of SR-HetGNN is more obvious.

\begin{figure}[htbp]
\centering
\includegraphics[scale=0.4,trim=260 110 260 110,clip ]{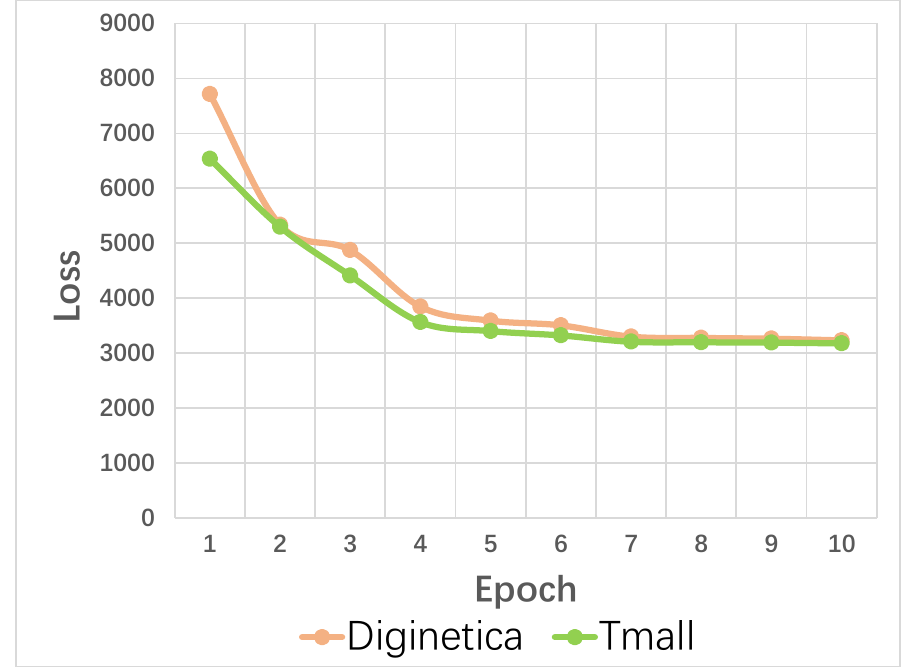}
\caption{Convergence process of the model. }
\label{fig:8}
\end{figure}

During the training process of the model, the objective function will get the optimal value at a certain moment. At this time, it is meaningless to continue training the model, because the performance of the model will hardly improve significantly. So it is important when to stop training the model to save time. In this paper, the loss curve of the model is drawn to judge when the model converges. The loss curve in the training process is shown in Fig. \ref{fig:8}.

As shown in Fig. \ref{fig:8}, the loss values decrease rapidly in the first four epochs, and the model is converged at the 10th epoch on Diginetica and Tmall dataset. Therefore, the number of the epoch can be set to 10 when training the model.

\subsection{Comparison with Baselines }
In order to test the performance of our model, it is compared with other recommendation methods. The results are shown in Table \ref{tableresult}.


\begin{table*}[htbp]
\centering
  \caption{The performance of SR-HetGNN with other baseline methods}
  \label{tableresult}
  \begin{tabular}{ccccc }
    \hline
    \multirow{2}{*}{Method} 
     &\multicolumn{2}{ |c| }{Diginetica} & \multicolumn{2}{ c }{Tmall} \\ 
    \cline{2-5}
     &\multicolumn{1}{ |c}{Recall@50} &\multicolumn{1}{c|}{MRR@50} & Recall@50 &MRR@50   \\ 
   
    \hline
    \ BPR-MF &  8.22 &  1.16  & 3.33  & 0.56 \\  
    \ FPMC   & 38.03 &	8.12  &	16.31 &	2.89  \\
    \ GRU4Rec& 35.11 &	9.53  &	15.72 &	5.67  \\
	\ SR-GNN & 62.91 &  18.98 & 24.34 & 7.13  \\
	\ TA-GNN & 62.49 &   19.36  & 23.77 & 5.89  \\
	\ SHARE & 59.93 &   19.57  & 17.41 & 4.65  \\
	\ GCE-GNN & 62.32 &  $\bm{22.03}$ & 20.52 & 4.06   \\
	\ MIHSG & $\bm{70.55}$  &   21.39 & 27.13 & $\bm{7.90}$  \\
    \hline
    \ SR-HetGNN& 64.24  & 18.15 & $\bm{28.03}$ & 7.27  \\
    \hline
    \end{tabular}
\end{table*}

It can be seen from the Table \ref{tableresult} that SR-HetGNN can achieve competitive results on both datasets, especially on the Tmall dataset. Because SR-HetGNN constructs the session sequence into a heterogeneous graph containing item nodes, user nodes, and session nodes. The heterogeneous graph contains complex dependencies, which can show the transitions between items, the connections between items and users. At the same time, SR-HetGNN can learn item embeddings by aggregating heterogeneous neighbors, which can mine more items related to the current session. However, the performance of SR-HetGNN in MRR@50 is not as good as that of TA-GNN on Diginetica. The possible reason is that TA-GNN employs a target-aware attentive
network to activate specific user interests with respect to a target item. SR-HetGNN only considers the sequence relations between items when constructing heterogeneous graphs and sampling heterogeneous neighbors. In the subsequent aggregation process, the sequential relationship between items is not further mined. It can be seen that SR-HetGNN is more suitable for the recall stage of the recommendation system than the sorting stage. SHARE can not achieve competitive results on the two data sets. This may be because hypergraph attention network is not suitable for recall tasks. Both GCE-GNN and SR-HetGNN networks make use of global information, but heterogeneous graph neural networks can capture more abundant interactive information, so HetGNN gets better results on the two datasets.SR-HetGNN does not perform as well as MIHSG on diginetica dataset. The reason may be that MIHSG can compress local context-sensitive user preferences from different granularity, improving recommendation performance. Compared with MIHSG,Het-SRGNN's competitive performance on the Tmall dataset shows its ability to model transaction sequences in real world, and Het-SRGNN has fewer parameters.

There are two traditional recommendation methods in the baseline methods adopted in this paper, BPR-MF, and FPMC. BPR-MF uses Matrix Factorization (MF) to learn user preferences, but MF does not work well on dealing with serialized item relationships, so its final results are not good; FPMC combines Matrix Factorization and Markov Chain (MC), where MF is used to learn user preferences, MC can construct the user's sequence behavior, but this recommendation method is difficult to learn the transfer of user preferences, that is user's short-term preferences. The performance of the FPMC model in the data set used in this paper is better than that of BPR-MF. The fundamental reason is the user sequence behavior established by the Markov Chain.

The core of GRU4Rec is GRU. GRU is a variant of Recurrent Neural Network (RNN). RNN is sensitive to serialized data and can learn the sequence patterns of user behaviors to produce better recommendations. However, the recommendation algorithm based on RNN can only simulate the one-way transfer between consecutive items, and can not consider the dependences of remote items, thus ignoring some information in the session sequence. SR-GNN constructs a homogeneous graph of items in session sequence, and it uses Graph Neural Network to learn item embeddings. The Graph Neural Network can learn the complex transformation relationship between items. It not only extracts the relationship between different items in the same session but also captures the relationship between items in different conversations, which makes the learning item embeddings more expressive. It can be seen from the experimental data in Table \ref{tableresult} that SR-GNN performs better than GRU4Rec. The reason is that the Graph Neural Network learns the complex transitions between items through the constructed graph. TA-GNN constructs target-aware embedding to adaptively consider the relevance of historical of behaviors concerning target items, it causes the advanced performance on MRR@50 of Diginetica. However, TA-GNN and SR-GNN are still limited. That is they all constructe the homogeneous graph based on the items in the session sequence. The homogeneous graph only has the transitions between items but ignores other information in the session sequence, especially user information.GCE-GNN can learn global information, which results in its excellent performance on MRR@50 of Diginetica.But the graph is full of item nodes, which cannot capture more complex and semantically rich information In general, SR-HetGNN can learn the user’s long-term preferences and short-term preferences, which cannot be done by the matrix factorization method that can only learn the user’s long-term preferences. Also, the Heterogeneous Graph Neural Network learns the complex dependencies between nodes in the Heterogeneous Graph, and its expressive ability in serialization is better than the Markov chain. 

\subsection{Ablation Experiment }
To further verify the effectiveness of different components of SR-HetGNN, this paper compares SR-HetGNN with the following  methods:
\begin{itemize}
\item The nodes of session type  are ignored when building the heterogeneous graph (SR-HetGNN-S).
\item The pre-embedding vectors generated by the deep walk algorithm are directly inputted to the attention layer. That is to say, the heterogeneous graph neural network is not used (SR HetGNN-Het).
\item Session embedding is generated by simple average of n item embeddings that constitute a session, which replaces attention network (SR HetGNN-Att).
\end{itemize}
The experimental results are shown in Table \ref{table5_3}. It can be seen from the table that the results of SR-hetGNN are the best, which indicates that these three components of SR-hetGNN have a positive impact on the recommendation results. Using session-type nodes enables the recommender system to capture the interdependence of items between different sessions. And attention network generates more expressive session embedding by combining global embedding and local embedding. At the same time, the heterogeneous graph neural network can aggregate the information of neighbor nodes and promote the development of the model.

\begin{table*}[htbp]
\centering
  \caption{Comparison of ablation experiments}
  \label{table5_3}
  \begin{tabular}{ccccc }
    \hline
    \multirow{2}{*}{Method} 
     &\multicolumn{2}{ |c| }{Diginetica} & \multicolumn{2}{ c }{Tmall} \\ 
    \cline{2-5}
     &\multicolumn{1}{ |c}{Recall@50} &\multicolumn{1}{c|}{MRR@50} & Recall@50 &MRR@50   \\ 
   
    \hline
    \ SR-HetGNN-Het& 61.72 & 17.49 & 27.38 & 6.36   \\  
    \ SR-HetGNN-S  & 62.28 & 17.59 & 27.51 & 6.49   \\
    \ SR-HetGNN-Att& 63.12 & 17.40 & 27.83 & 6.98  \\
	\hline
    \ SR-HetGNN& $\bm{64.24}$ & $\bm{18.15}$ & $\bm{28.03}$ &$\bm{7.27}$  \\
    \hline
    \end{tabular}
\end{table*}

\subsection{Case Study}

To verify our model's ability to understand conversational context and learn complex transformations between items, we extract specific sessions from the test set of diginitica for analysis. These sessions are grouped based on their shared target item, resulting in the identification of the ten groups with the highest session count. Subsequently, these sessions are employed as input for the SRGNN and SR-HetGNN models, enabling the derivation of session vectors. Subsequent t-SNE analysis is conducted on these vectors to visualize the relationships and patterns among the conversations within each group.The statistics of extracted session groups are shown in Table \ref{table6_1}.

\begin{table}[h]\small
\centering
  \caption{Statistics of extracted session groups}
  \label{table6_1}
  \begin{tabular}{cl}
    \hline
    Target item& Count\\
    \hline
    12313&   11\\
    12270&   9\\
    14246&   8\\
    10377 &   8\\
    13463 &   8 \\
    10209 &   8\\
    1340&   8\\
    2244 &   7\\
    9268 &   7\\
    \hline
    \end{tabular}
\end{table}

\begin{figure}[H]
\centering
\includegraphics[scale=0.35,clip ]{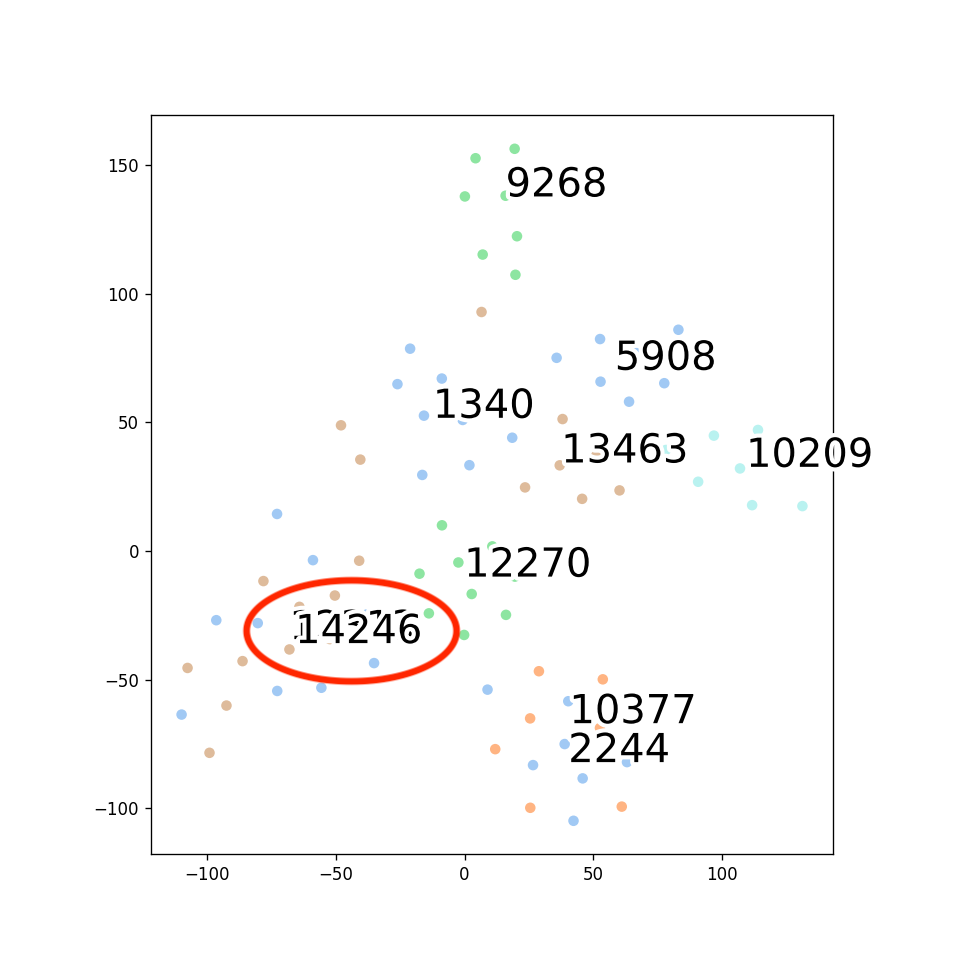}
\caption{SR-HetGNN on diginetica. }
\label{fig:9}
\end{figure}

\begin{figure}[H]
\centering
\includegraphics[scale=0.35,clip ]{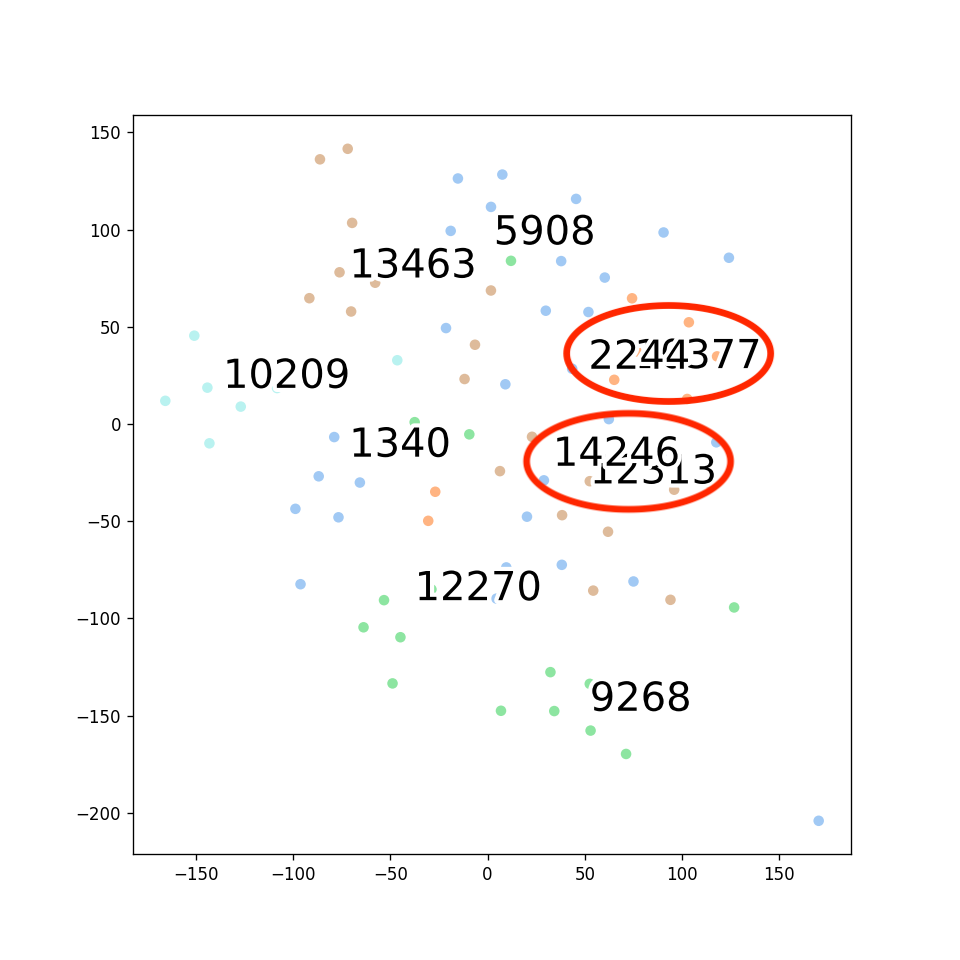}
\caption{SRGNN on diginetica. }
\label{fig:10}
\end{figure}

As shown in Fig. \ref{fig:9}. and Fig. \ref{fig:10}., valuable insights can be obtained through the visual examination of the resulting scattered point distribution. It is noteworthy that our model successfully constrains sessions with the same target item to adjacent spaces, with fewer overlapping session vectors distributed. This observation serves as evidence that our model effectively incorporates session context and demonstrates enhanced learning capabilities in handling complex transitions between items.

\section{Conclusion}\label{sec6}

This paper proposed SR-HetGNN, a session-based recommendation method that applies Heterogeneous Graph Neural Network to learn item and session embeddings. By constructing a heterogeneous graph with multiple types of nodes and utilizing the Attentional Network to generate session embeddings, SR-HetGNN captures complex transformation relationships between items and user preferences from session sequences. The experiments demonstrated that SR-HetGNN outperformed other popular methods.

Although SR-HetGNN has shown promising results, the current dataset limitations only allowed for three types of nodes to be considered when constructing the heterogeneous graph. Future work will focus on expanding to real-world datasets with multiple types of nodes, increasing the learned session embeddings' expressiveness. Furthermore, sorting optimization will be incorporated into the model to further improve the ranking of recommended results.
\backmatter

\bmhead{Acknowledgments}

This work was supported in part by Beijing Natural Science Foundation(Grant No.L233034), in part by Zhejiang Lab Open Research Project (Grant No.K2022KG0AB03), in part by the Open Projects of the Technology Innovation Center of Cultural Tourism Big Data of Hebei Province (Grant No.SG2019036-zd202205), in part by the National Natural Science Foundation of China (Grant No.62002027, No.72274022), in part by CCF-Zhipu AI Large Model Fund (Grant No. CCF-Zhipu202317) and in part by the Fundamental Research Funds for the Central Universities (Grant No.2023RC08, No.21623402).





\bibliographystyle{sn-mathphys}
\bibliography{mybibfile}


\end{document}